\documentclass[11pt]{article}
\usepackage{graphics}
\usepackage{graphicx}
\usepackage{amssymb}
\usepackage{amsmath}
\usepackage{multirow}
\usepackage{caption}
\usepackage{subcaption}
\usepackage[table]{xcolor}
\usepackage[utf8x]{inputenc}

\title{How connected is too connected? Impact of network topology on systemic risk and collapse of complex economic systems}

\date{}
\usepackage{authblk}
\makeatletter
\renewcommand\AB@affilsepx{. \protect\Affilfont}
\makeatother
\author[1,3,4]{Aymeric Vi\'{e}}
\author[1,2,*]{Alfredo~J. Morales}
\affil[1]{New England Complex Systems Institute, 277 Broadway, Cambridge, MA, 02143, USA}
\affil[2]{MIT Media Lab, USA}
\affil[3]{Paris School of Economics, France}
\affil[4]{Sciences Po Saint-Germain-en-Laye, France}
\affil[*]{corresponding author: alfredo@necsi.edu}

\begin{document}


\maketitle
 \begin{abstract}
Economic interdependencies have become increasingly present in globalized production, financial and trade systems. While establishing interdependencies among economic agents is crucial for the production of complex products, they may also increase systemic risks due to failure propagation. It is crucial to identify how network connectivity impacts both the emergent production and risk of collapse of economic systems. In this paper we propose a model to study the effects of network structure on the behavior of economic systems by varying the density and centralization of connections among agents. The complexity of production increases with connectivity given the combinatorial explosion of parts and products. Emergent systemic risks arise when interconnections increase vulnerabilities. Our results suggest a universal description of economic collapse given in the emergence of tipping points and phase transitions in the relationship between network structure and risk of individual failure. This relationship seems to follow a sigmoidal form in the case of increasingly denser or centralized networks. The model sheds new light on the relevance of policies for the growth of economic complexity, and highlights the trade-off between increasing the potential production of the system and its robustness to collapse. We discuss the policy implications of intervening in the organization of interconnections and system features, and stress how different network structures and node characteristics suggest different directions in order to promote complex and robust economic systems.    

 \end{abstract}


\newpage

\section{Introduction}

\subsection{Production networks and risk}

As highly complex and networked systems, the properties of economies are characterized by the behavior and interdependencies of their components (Bar-Yam, 1997; Hidalgo et al., 2007; Barrat et al., 2008). Whether they arise from investments, trade or supply chains, interdependencies are increasingly important in contemporary economic systems, and fundamental for risk assessment and evaluation (Schewitzer et al., 2009).
Interconnections enable the diversification of output, improve the efficiency of economies, and increase the growth of economic complexity through the product space (Hidalgo et al., 2007). At the same time, they also introduce paths for risk contagion and generate large-scale vulnerabilities to systemic failure (Bar-Yam, 2010; Harmon et al., 2010). Given the current context of increasing international trade, financialization and globalization of economies, it is crucial to understand the effects of connectivity on networked economies and its relationship to economic collapse.

Traditional economic studies focus on explaining collapse through the contribution of different factors such as bankruptcy (Battiston et al., 2007), bank loans (Stiglitz and Greenwald, 2003), interbank credits (Allen and Gale, 2000), and changes of asset prices (Kiyotaki and Moore, 1997). Roukny et al. (2018) show that interconnections in bank systems through credit contracts and subject to correlated external shocks constitute a source of uncertainty in systemic risk assessment. 
Additional studies have indeed emphasized on the need for understanding the impact of network structure on the probability of collapse (Schweitzer et al., 2009; Iyer et al., 2013; Albert et al., 2000). Battiston et al. (2012b) emphasized their study on the identification of important nodes through feedback centrality with debt ranking. These studies show an inherent relationship between the structure of the network and its robustness and vulnerability to selected attacks and random errors independently of their nature. Stressing the systemic complexity of economic networks may contribute to the design and implementation of policies leading to higher diversity and efficiency without undermining the robustness of economic systems (Schwitzer et al., 2009; Battiston et al., 2012b). 

In this paper we develop a model to show that while the creation of interdependencies among economic agents is fundamental for the growth of economic complexity, it also amplifies the risk of collapse during adverse conditions. We show that the structure of interconnections among economic agents increases the fragility of economic systems despite an apparent improvement of their production complexity. We explore two different ways in which systems can be interconnected: density and centralization. Density refers to the number of connections that are drawn among agents independently. Centralization refers to the emergence of highly connected nodes that bridge across large parts of the network. While different in nature, these directions show how nodes can become increasingly dependent on one another, either directly or indirectly (through secondary connections). We found that the transition to collapse is universal and independent of a specific network structure. Instead the transition results from the reachability of nodes to one another and the spreadability of their failure.

The architecture of economic networks is crucial to study their efficiency and vulnerabilities to systemic failure. For example, if a vital firm within a supply chain suddenly ceases to exist
, all producers linked to the failing element become unable to produce their output. The present analysis aims at quantifying both robustness and performance of production networks for various levels of individual node failure, and for various degrees of density and centralization in the network. 
Interconnections may under certain conditions increase the complexity of production. However, as we introduce a non-zero probability of failure, the expected diversity and productivity decreases, leading the way to economic collapse.

There is a large literature on network robustness to internal failure and external attacks on nodes or edges. Previous studies have mainly focused on two particular network topologies: the Erdos-Rényi random graph (Erdos and Rényi, 1960) and Barabási's scale-free network (Barabási et al., 1999; Barabási and Bonabeau, 2003). Albert, Jeong and Barabási (2000) studied error tolerance and attack impact, notably testing both web robustness to targeted attacks on well connected nodes, and to removal of a given nodes fraction. Crucitti et al. (2003) study network robustness to failure and targeted attacks. Iyer et al. (2013) likewise analyze how interconnections structure evolve with the removal of vertices, for a variety of networks types. Lorenz, Battiston and Schweitzer (2009) developed a general framework to systemic risk with cascading failures processes in networks through node fragility. Pichler et al. (2018) investigated the issue of systemic risk in the context of efficient asset allocation in the form of a network optimization problem. Caccioli et al. (2018) recently provided a thorough review of research in network models of financial systemic risk. Buldyrev et al. (2010) extended the framework of network cascading failure analysis to the case of interconnected networks transmitting failure from one to another. In the spirit of Albert et al. (2000) who focused on two models: the Erdos-Rényi random network model (Erdos and Rényi, 1960) and the scale-free web (Barabási et al., 1999; Barabási and Bonabeau, 2003), the present paper extends the investigation on the robustness of networks models to a more general framework, including the transition from random to scale-free networks and further centralization, as well as the effects of density of connections.

\subsection{Production and collapse}

Various indicators of network robustness, and by consequence its fragility, have been extensively used in the literature. The communication capacity of the network after nodes removal has been analyzed by Crucitti et al. (2003). Albert et al. (2000) used the average shortest path length among nodes in the network as a indicator of failure reachability. Iyer et al. (2013) and Callaway et al. (2000) analyzed robustness as percolation efficiency on networks. Lorenz et al. (2009) used the fraction of stable nodes after removing the failed ones as a measure of systemic risk. Rather than providing an indicator of fragility, we show the space of possible behaviors of networked production systems in terms of diversity of outcome and probability of collapse for different scenarios regarding conditions to failure and structure of interdependencies. 
%
%
%


Collapse may be framed as a comparison of the current state of the system with respect to a reference one. We define the reference state as the situation of autarky or network-free environment, where agents have no interaction with each other.  A production below such reference state could be considered as collapse, i.e. a systemic failure of the network to achieve the network-free production levels. Because the reference state is defined without knowledge of the networked structure of the system, collapse is interpreted in our model as the inability of the system to achieve the autarky production level. Our results are generalizable and consistent with other definitions of collapse.

This article is organized as follows. Section 2 describes the network generation algorithms and the production model. Section 3 presents the results of model simulations in terms of productivity and collapse probability across multiple network topologies. Their discussion, implications and relations to previous literature are provided in section 4. Section 5 concludes on the impact of network connectivity structure on global production and risk of failure in economic systems.

\section{Model}

We design a simple economic model of production, structured by a network of partnerships or supply chains. Nodes are represented as economic agents, such as individuals, firms or countries. Links indicate economic interdependencies. In order to produce goods, networked agents need the input from their connections. Each node has an individual error probability, analogous to the possibility of node removal in previous literature. Errors propagate through cascades across the network. We consider two network generation processes respectively based on the density or centralization of connections. In this section we present the network generation processes, and define the mechanisms for production and collapse. 

\subsection{The density network model}
\label{subsection_density_network}

\begin{figure}[!tb]
\minipage{0.32\textwidth}
  \includegraphics[width=\linewidth]{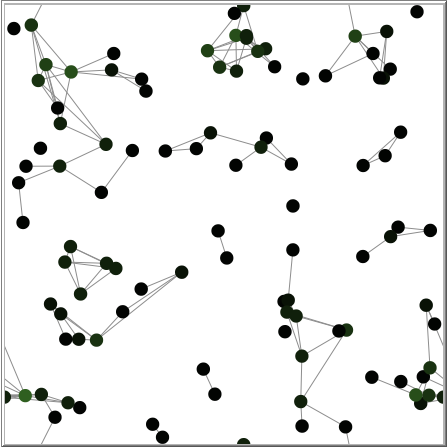}
\endminipage\hfill
\minipage{0.32\textwidth}
  \includegraphics[width=\linewidth]{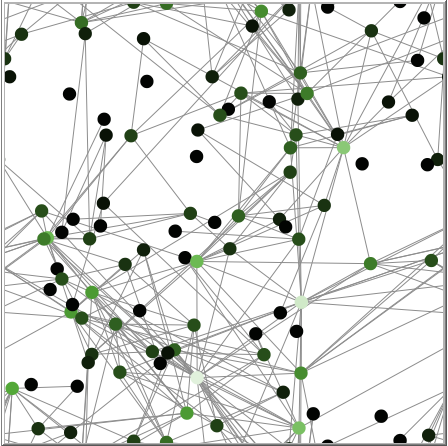}
\endminipage\hfill
\minipage{0.32\textwidth}%
  \includegraphics[width=\linewidth]{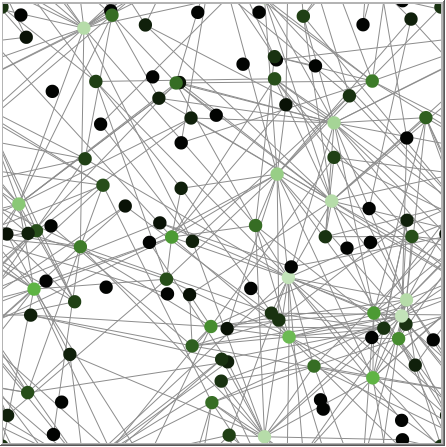}
\endminipage
\caption{Visualization of network topologies from the density model. Panels show networks that result from different model parameters. The density of the networks increase from left to right. The node color is proportional to the degree (from black to green). Reach $r= 4$ and density $\delta= 0.028$ in the left panel. Reach $r= 10$ and density $\delta= 0.053$ in the middle panel. Reach $r= 16$ and density $\delta= 0.058$ in the right panel.}
\label{density_fig}
\end{figure}

In order to analyze the impact of network density over production and risk of failure, we create a network generation model. Nodes are randomly distributed in a torus space. Their connections depend on their distance to each other and a threshold denoted {\it reach}. Nodes first create a link with a randomly chosen node within the reach distance, denoted {\it target}, and second create links with all nodes linked to the target node. The probability of a node $i$ to initially create a link to a node $j$ at distance $x_{ij}$ is as follows:

\begin{equation}
p_{ij}(x_{ij}) =
\begin{cases} 
   \frac{1}{N(r)} & \text{if } x_{ij} \leq r \\
   0              & \text{if } x_{ij} > r
\end{cases}
\end{equation}

where $r$ indicates the reach radius and $N(r)$ represents the total number of nodes at reach from $i$. The number of potential target nodes $N(r)$ increases proportionally to the reach parameter $r$. The density of the network ($\delta$) indicates the ratio between the number of existing edges divided by the total number of possible edges in the network. Low reach values yield only a few connections. As the reach parameter increases, so does the number of connections and the network density. Figure \ref{density_fig} illustrates networks that result from different model parameters. The density of the networks increase from left to right. 


\subsection{The centralized network model}
\label{subsection_centrality_network}

The centralized network model generates graphs with different levels of centralization. For this purpose, we generalize the preferential attachment mechanism (Barab\'{a}si and Albert, 1999) with an exponent that controls for the emergence and importance of hubs-- ranging from no centralization (independently distributed edges) to perfectly centralized networks (in one or a couple hubs). In between the two extreme cases, we obtain a wide range of scale-free networks where several hubs are present with different relative importance in the graph.

\begin{figure}[!tb]
\minipage{0.32\textwidth}
  \includegraphics[width=\linewidth]{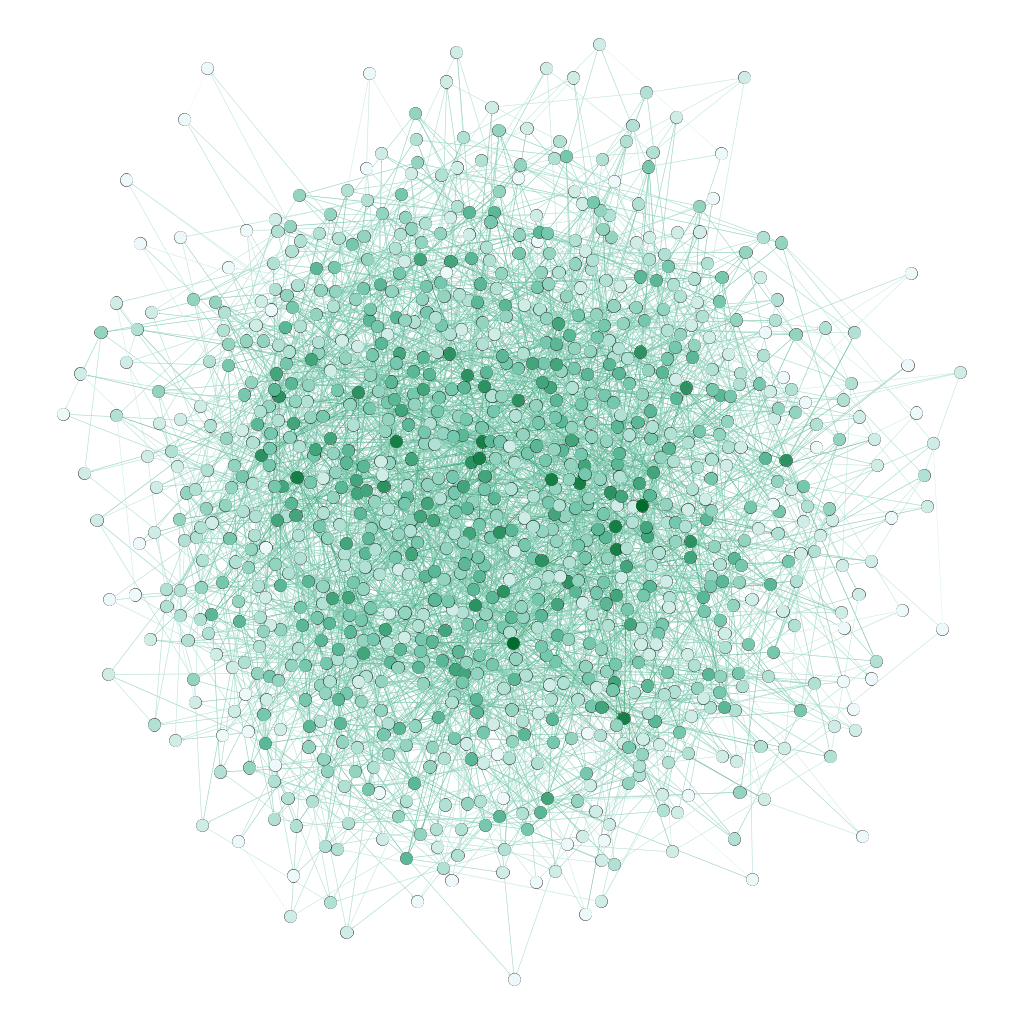}
  \caption*{$\alpha = 0$}\label{alpha0}
\endminipage\hfill
\minipage{0.32\textwidth}
  \includegraphics[width=\linewidth]{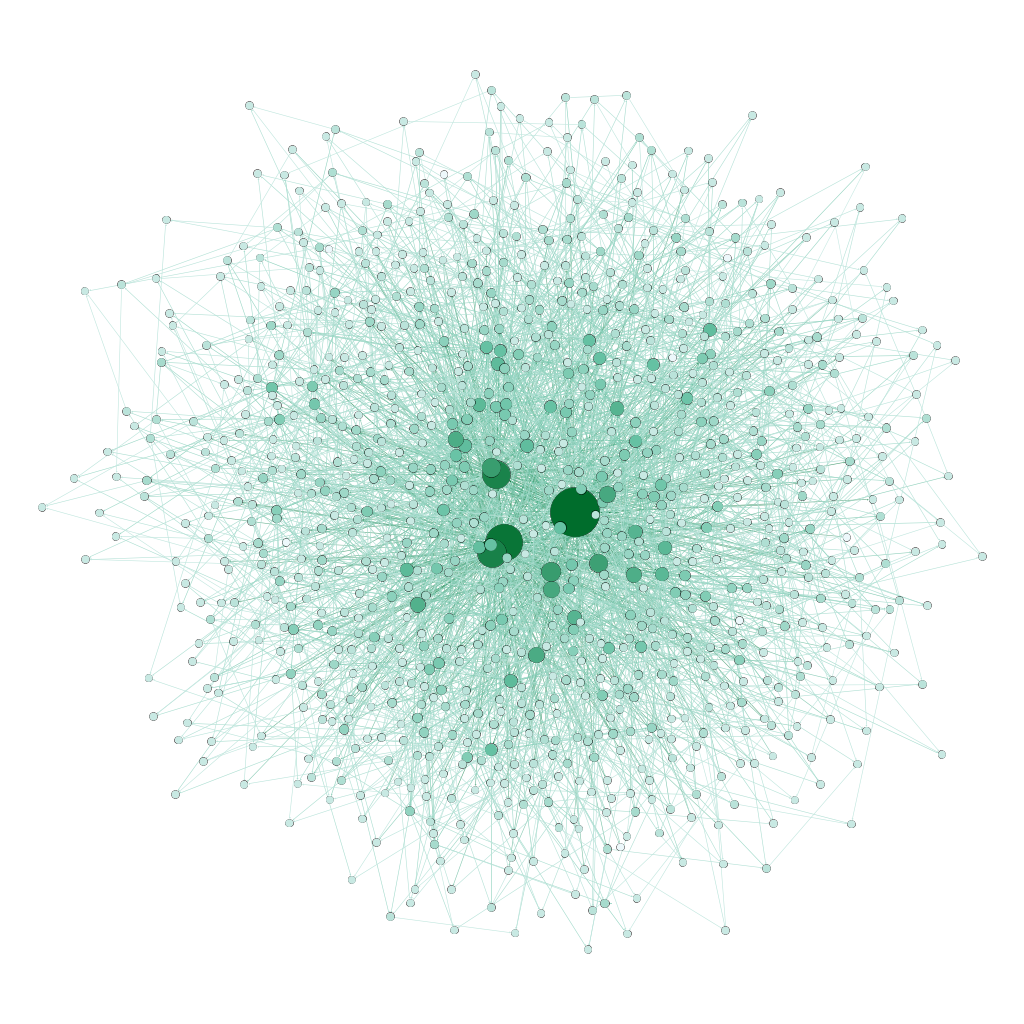}
  \caption*{$\alpha = 1$}\label{alpha1}
\endminipage\hfill
\minipage{0.32\textwidth}%
  \includegraphics[width=\linewidth]{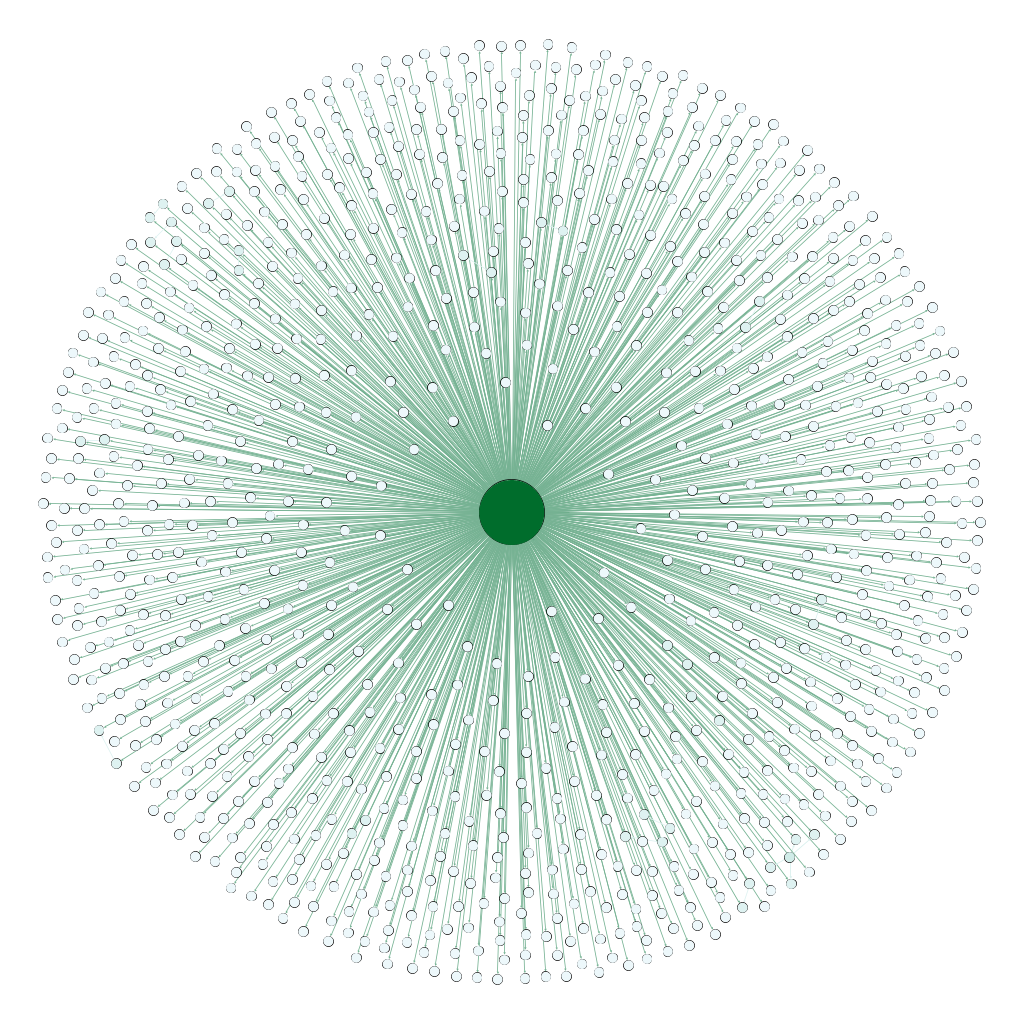}
  \caption*{$\alpha = 2$}\label{alpha2}
\endminipage
\caption{Visualization of network topologies from the centralized model. Panels show networks that result from different model parameters. From left to right the centralization of the network increases. The node color is proportional to the degree (from white to green). The left panel shows a decentralized network, similar to the Erdos-Rényi model ($\alpha=0$). The middle panel shows a scale-free network, similar to the Barabasi-Albert model ($\alpha=1$). The right panel shows a perfectly centralized network ($\alpha=2$). The number of edges and density is constant across all networks ($\delta= 0.02$).}
\label{alpha_fig}
\end{figure}

The network generation process consists in creating edges as a function of the attachment probability. The probability of node $i$ to create an edge with node $j$ is as follows:

\begin{equation}
    p_{ij} \propto k_j ^ \alpha
\end{equation}

where $k_j$ is the number of connections of node $j$ and $\alpha$ is the exponent we use to control the influence of the preferential attachment mechanism. If $\alpha = 0$, the attachment probability becomes equal among all nodes and we obtain a random network with no central hub similar to the Erdos-Ranyi model. If $\alpha = 1$, we obtain the standard Barab\'{a}si-Albert network with a few hubs. If $\alpha = 2$, we create a network with full centralization where all nodes are linked to a single central one. This extension of the preferential attachment mechanism magnifies the degree heterogeneity among nodes for $\alpha > 1$, and reduces such attractive force for any $\alpha < 1$. An illustration of the model variants is shown in Figure \ref{alpha_fig}. 






\subsection{Production and collapse}

The production mechanism is invariant across both network generation methods. In order to run simulations, we create $100$ economic agents and interconnect them following the steps described in sections \ref{subsection_density_network} and \ref{subsection_centrality_network} respectively. Once the networks are created, we simulate both production and failure. 
 
Agents produce goods with uniform and constant production technology. 
We consider production inputs as received endowments, without introducing stock constraints or resource extraction. Agents produce as many goods as possible, under the constrains imposed by the piece-wise production function described as follows: 

\begin{equation}
\label{production_function}
    q_i = 
    \begin{cases}
       1 & \text{if } k_i = 0 \\
       2n & \text{if } k_i = n, n \in \mathbb{N}_{>0}
    \end{cases}
\end{equation}

\noindent  where $q_i$ denotes production of node $i$, and $k_i$ its degree (number of connections). A node without connections ($k_i = 0$) will only produce $1$ good. We define this state as autarky. A node of $n$ connections will be able to produce $2n$ goods (assuming no failure). The hypothesis behind introducing a production scaling parameter is derived from the view of production as a combinatorial process (Hidalgo et al., 2007). Economically speaking, it may account for increasing returns to scale, heterogeneity in marginal cost or differences in production efficiency. 

Agents have an identical and exogenous failure probability $p$. It conveys the individual probability of encountering issues in the production process and not providing any output at a given period. This modelling specification can be related to error tolerance (Albert, Jeaong and Barabási, 2000) and removal probability (Crucitti et al., 2003). We define this phenomenon as individual failure. Individual failure may happen due to resource shortage, production tools dysfunction, or any other exogenous event leading to null production. Global failure arises as individual failures cascade across the network. Individual failure is denoted {\it failure probability}, while global failure is denoted {\it collapse probability}. 

Individual failure spreads across the network through direct connections, i.e. to the economic partners directly linked to the failing node. We do not spread failure to neighbors of neighbors in this simple contagion mechanism. 
Our node-driven approach is closely related to Battiston et al. (2007, 2012a) who start from local interactions to study systemic failure, providing a new framework for understanding failure propagation. Our direct propagation framework enables the analysis of cascading failures, which is an important phenomenon to be considered in the study of network robustness (Crucitti et al., 2013). 

Accounting for failure probability $p$, and with $k_i$ nodes directly linked to node $i$, the piece-wise conditional expected production function $E(q_i|p)$ can be defined as follows:

\begin{equation}
\label{average_production}
    E(q_i|p) = 
    \begin{cases}
       (1 -p) & \text{if } k_i = 0 \\
       2n(1-p)^n & \text{if } k_i = n, n \in \mathbb{N}_{>0}
    \end{cases}
\end{equation}

If $k_i = n > 0$, node $i$'s production is equal to $2n$ with probability $(1-p)^n$ and equal to 0 with probability $1- (1-p)^n$. A sub network of $n$ directly connected agents has a collective probability $(1-p)^n$ of not failing (i.e. all nodes produce). With probability $1 - (1-p)^n$, at least one node fails and the entire sub network is not able to produce. Given that the production scales by a factor of $2n$, the expected production function takes the value $2n(1-p)^n$ for any $n > 0$. For an autarkic node $i$, for which the number of neighbors $k_i$ is null, the productivity is equal to 1, adjusted to the probability $p$ of failing at each experiment. The expected production at each experiment for such node is thus equal to $(1 -p)$.

This production function specification illustrates the trade-off we examine between inter-connectivity and risk. In the density model, higher connectivity results in both better possible production, but also increased risk on the entire supply chains. In the centralized network, the central node has the potential to deliver a huge output, but is vulnerable to the failure of any other node it is connected to. The indicators developed in the next subsection allow us to measure these phenomena.

\subsection{Measuring collapse and efficiency}
\label{collapse_definition}

The model is run for a given number of independent periods or experiments. We consider the failure and contagion processes as being transient. The failure of a node at a given experiment does not affect its state on the next experiment. This choice of simplicity identifies the impact of the network structure over systemic risk and productivity. 
 We define the system's total production $T$ at each experiment $t$ as the sum of the individual agent production levels $q_i$ as follows:

\begin{equation}
    T_t = \sum_{i = 1}^Nq_i
\end{equation}

\noindent where $N$ is the total number of nodes in the network. 

At a given experiment, we define {\it collapse} as the situation where the total production of the system $T_t$ is below a reference level $\lambda$, which is the expected production of the system in the autarky regime. 
 
 \begin{equation}
    \lambda(N,p) = N(1 - p) 
 \end{equation}
 
 The production  level  $\lambda(N,p)$ does not depend on the network structure. It enables the performance analysis of any network topology with respect to the autarky case, and evaluate whether any particular system architecture is expected to yield higher or lower production. Through numerical simulations of the model, we apply Monte-Carlo to estimate the collapse probability over 10.000 independent experiments for each possible set of the model parameters, including different levels of failure probability and network topologies.
 
 The model is implemented in the agent-based environment Netlogo (Wilensky, 1999). We use the Pattern Space Exploration and Sampling algorithms from the OpenMole platform (Ch\'{e}rel et al., 2015, Reuillon et al. 2010, 2013) to identify areas of high variation in the model results and improve both tractability and validation. More details about the model implementation and simulation methodology can be found in the Supplement (Section \ref{implementation}). 

\section{Results}

\subsection{Mapping system productivity}

\begin{figure}
\centering
\begin{subfigure}{.5\textwidth}
  \centering
  \includegraphics[width=1.00\linewidth]{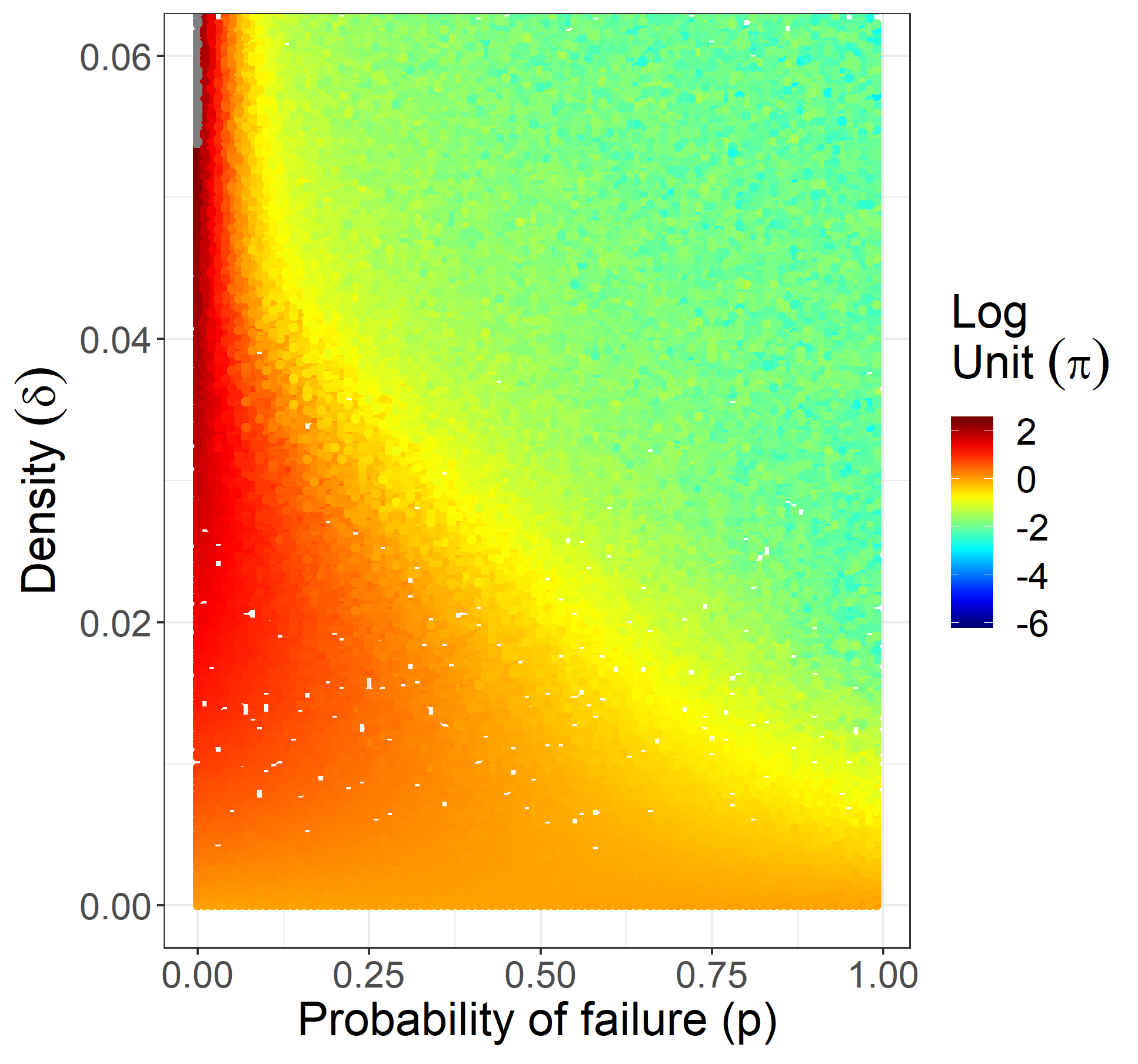}
  \caption{Density model}
  \label{productivity_density}
\end{subfigure}%
\begin{subfigure}{.5\textwidth}
  \centering
  \includegraphics[width=1.00\linewidth]{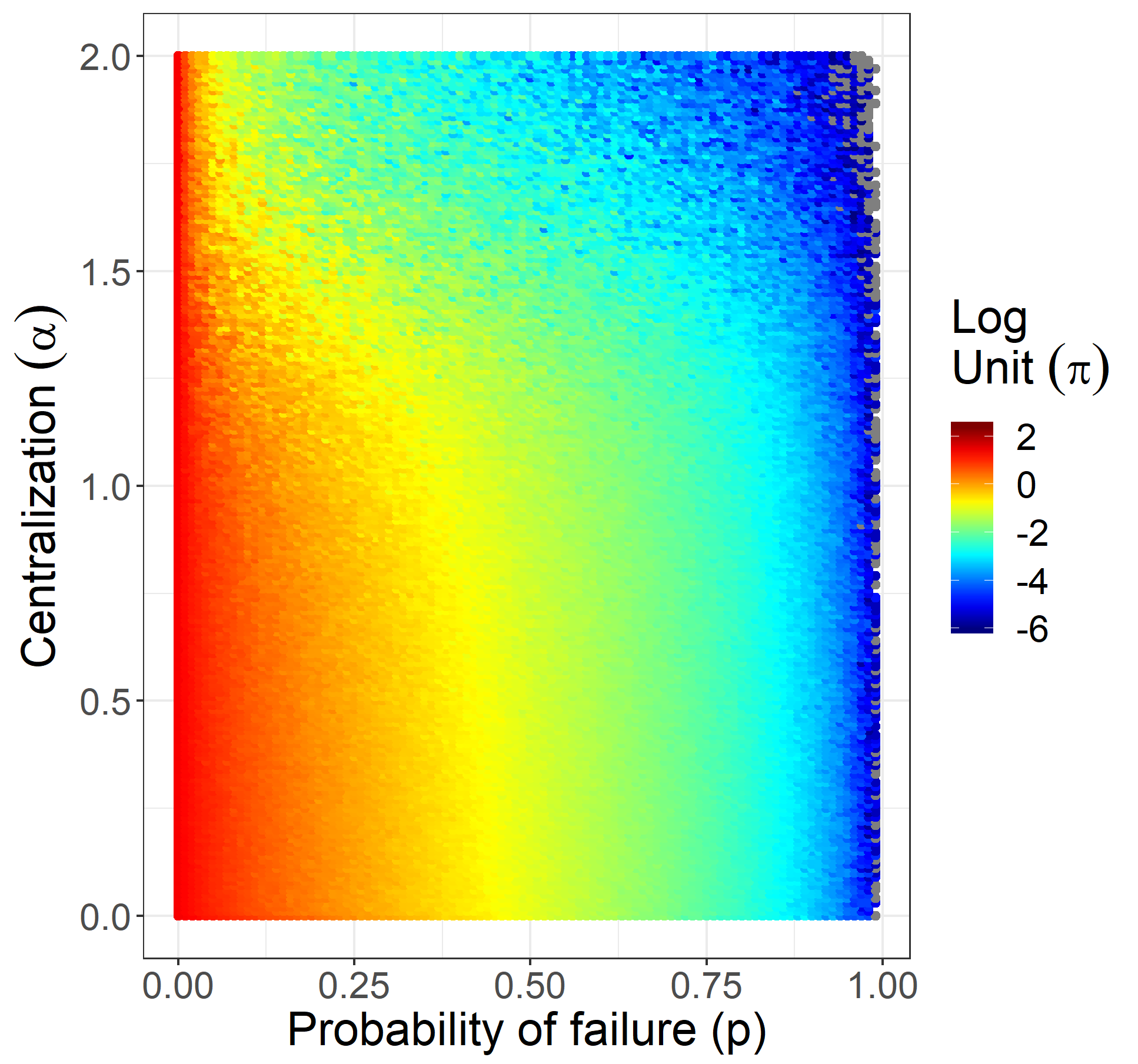}
  \caption{Centralized model}
  \label{productivity_centrality}
\end{subfigure}
\caption{Network productivity as a function of model parameters. 
Color indicates average productivity (log unit) in units of the autarky level on logarithmic scale. The left panel shows the outcomes of the density model. The right panel shows the outcomes of the centralized model. The x-axis represents the probability of individual failure in both panels. The y-axis represents the network density (left panel) or centralization (right panel). Scale shown in figure.}
\label{productivity_sampling}
\end{figure}


We define the system productivity level $\pi$ as the ratio of the system average production (equation \ref{average_production}) and the reference autarky production level. A production level of $\pi=2$ indicates that the system is able to double the autarky output level. Figure \ref{productivity_sampling} shows the production levels (colored regions) of different network structures as a function of individual probability of failure (x-axis) and the parameters  (y-axis) of the density (left) and centralized (right) models respectively. Red regions in Figure \ref{productivity_sampling} indicate high production, and green and blue regions indicate lower production. In the Supplement (subsection \ref{pse_figures}) we provide additional figures from the Pattern Space Exploration procedure used to determine areas of variation.

In both models, increasing the density or centralization of network connections results in higher production levels when the probability of failure is low (red regions near the vertical axis), given the possibility of agents to establish interdependencies and combine elements to create more complex products. However, as the probability of failure increases, the  average output decreases with the density or centralization of connections. This effect is more abrupt in centralized systems (right panel). Therefore, increasing the number of interdependencies may increase the complexity of the economic systems but it also makes them more fragile to individuals' failure.
 


\subsection{Mapping the collapse probability}

\begin{figure}
\centering
\begin{subfigure}{.5\textwidth}
  \centering
  \includegraphics[width=1.03\linewidth]{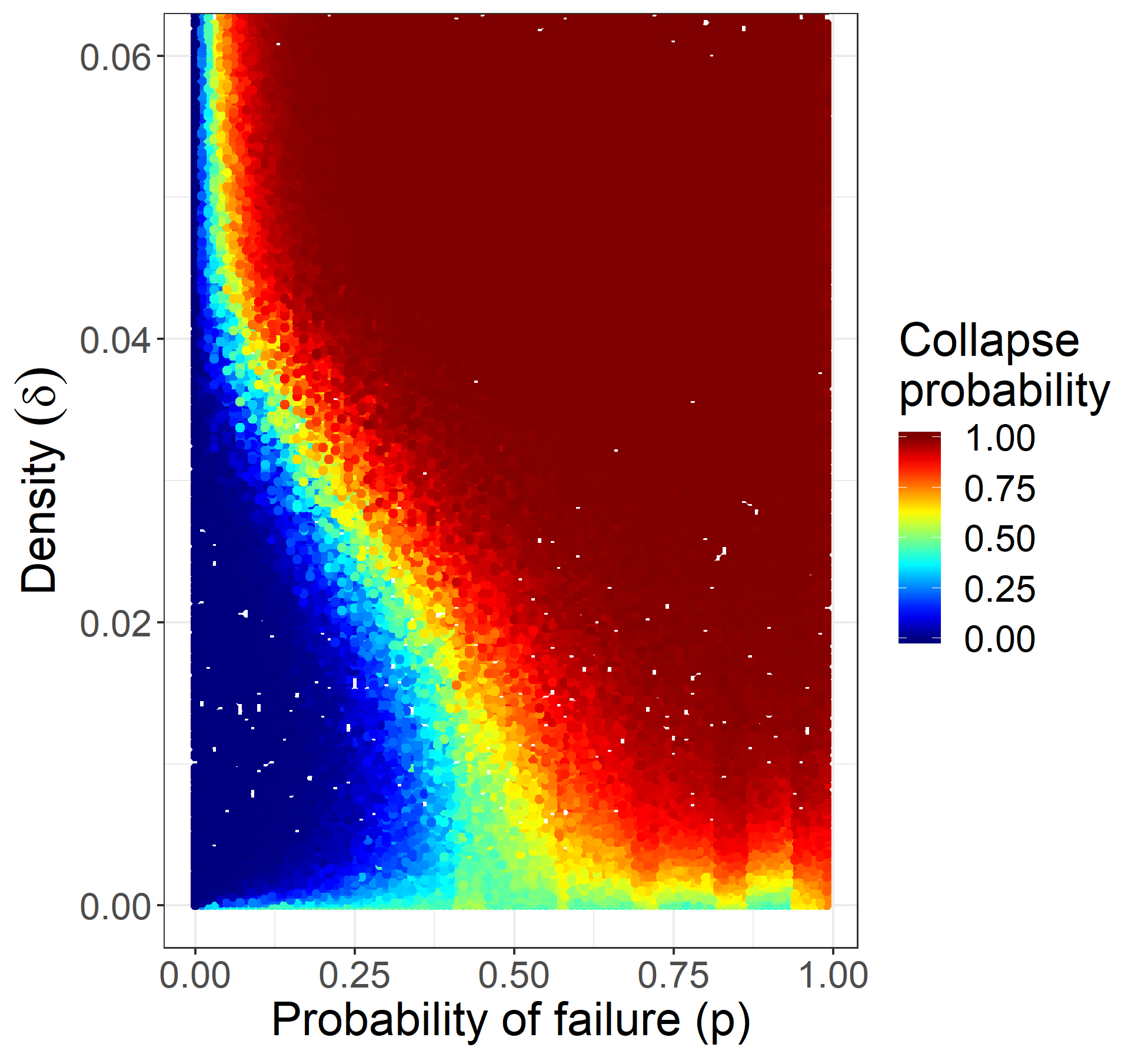}
  \caption{Density model}
  \label{density_proba}
\end{subfigure}%
\begin{subfigure}{.5\textwidth}
  \centering
  \includegraphics[width=1.03\linewidth]{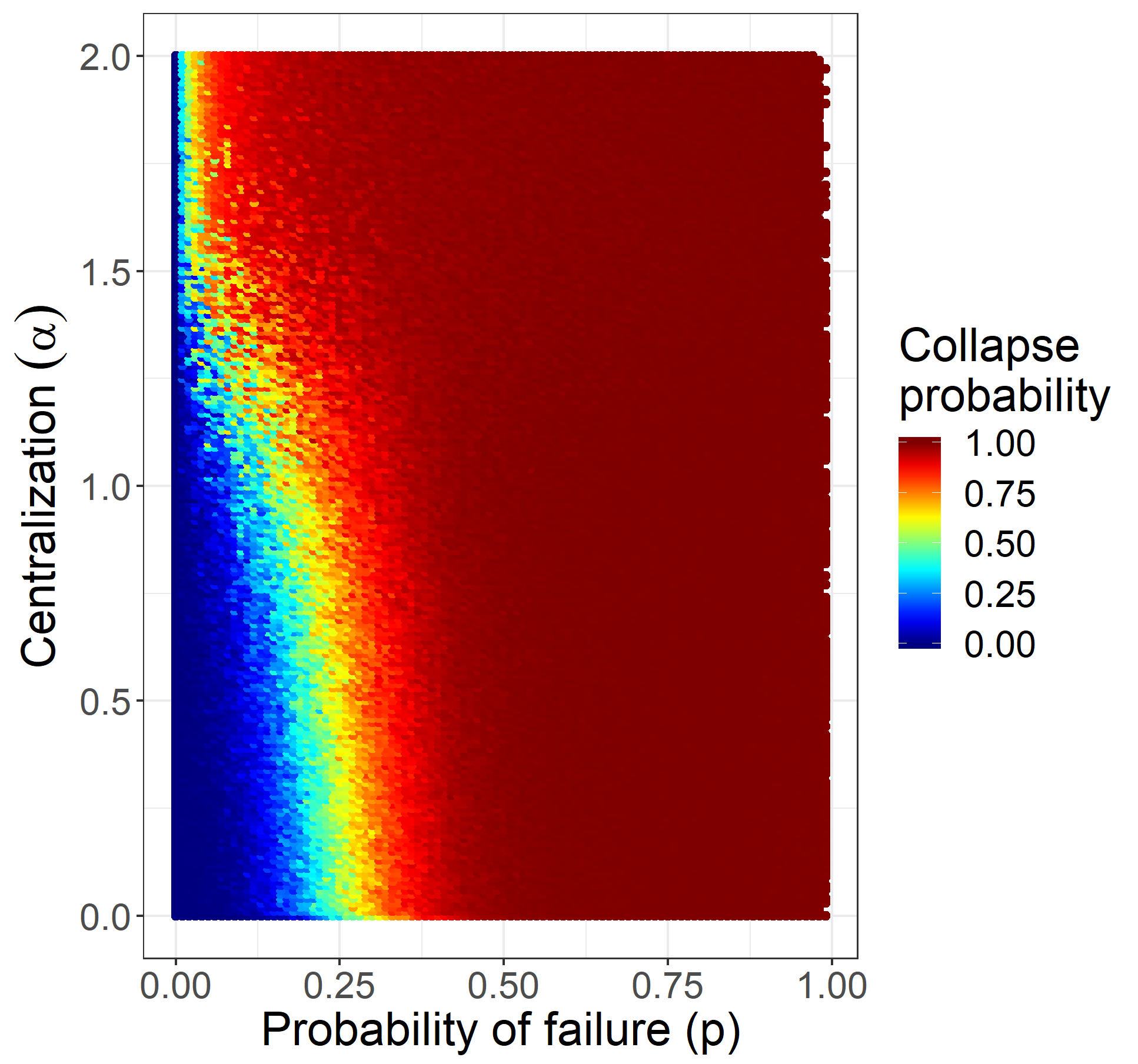}
  \caption{Centralized model}
  \label{centrality_proba}
\end{subfigure}
\caption{Probability of systemic collapse as a function of model parameters. Color indicates the collapse probability. The left panel shows the outcomes of the density model. The right panel shows the outcomes of the centralized model. The x-axis represents the probability of individual failure in both panels. The y-axis represents the network density (left panel) or centralization (right panel). Scale shown in figure.}
\label{collapse_sampling}
\end{figure}


%

While interconnections enable the creation of economic complexity, they also increase the probability of failure propagation and consequently the risk of global collapse. Figure \ref{collapse_sampling} provides a precise mapping of system collapse probability as a function of network structure (given in the parameters of the network density and centralization models) and probability of failure. Blue regions indicate very low risk of collapse. Red areas show very high risk of collapse. In both models, there is a region where the probability of collapse is low (blue). In these regions the productivity of the system is also high, as we previously noticed in Figure \ref{productivity_sampling}. The probability of collapse increases when we either increase the probability of failure, for a given network setup, or when we increase the number of interdependencies, for probabilities of failure that are not close to zero. 

The transition from robust (blue) to fragile (red) systems seem to be very sharp (yellow and green regions in Figure \ref{collapse_sampling}). This indicates the existence of tipping points for each network setup. Moreover, the location of the tipping point changes as we modify the network structure or failure probability. It comes closer to the vertical axis as we increase the density or centralization of the network. Denser or more centralized networks are more sensitive (or fragile) to individual failure. 

In Figure \ref{productivity_sampling} we showed that the highest levels of production take place when the density or centralization of connections is highest and the probability of failure is lowest (upper left corner). In Figure \ref{collapse_sampling} we notice that such region is also the most fragile to an increase of the probability of failure. Notice that the number of links is constant in the centralized model and only the centralization of edges around hubs changes as we increase the parameter $\alpha$. Therefore, two radically different network models present remarkable similarities in their behavior, which shows that centralizing interdependencies in a few nodes is just as potentially harmful as creating an excess of them in a distributed manner.



\subsection{Productivity and network structure}

\begin{figure}[t!]
\centering
\begin{subfigure}{.5\textwidth}
  \centering
  \includegraphics[width=1.0\linewidth]{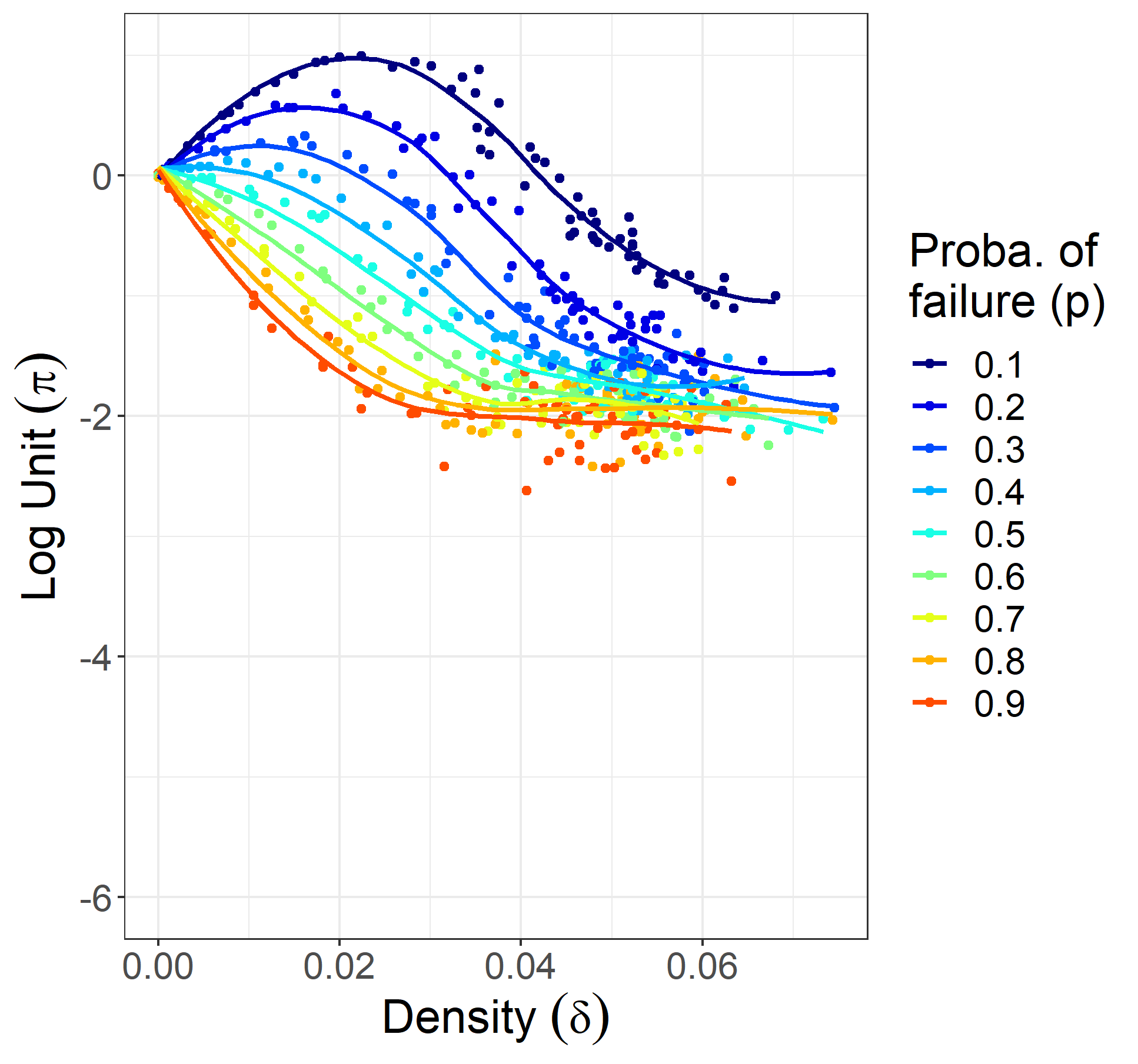}
  \caption{Density model}
  \label{productivity_density_transitions}
\end{subfigure}%
\begin{subfigure}{.5\textwidth}
  \centering
  \includegraphics[width=1.0\linewidth]{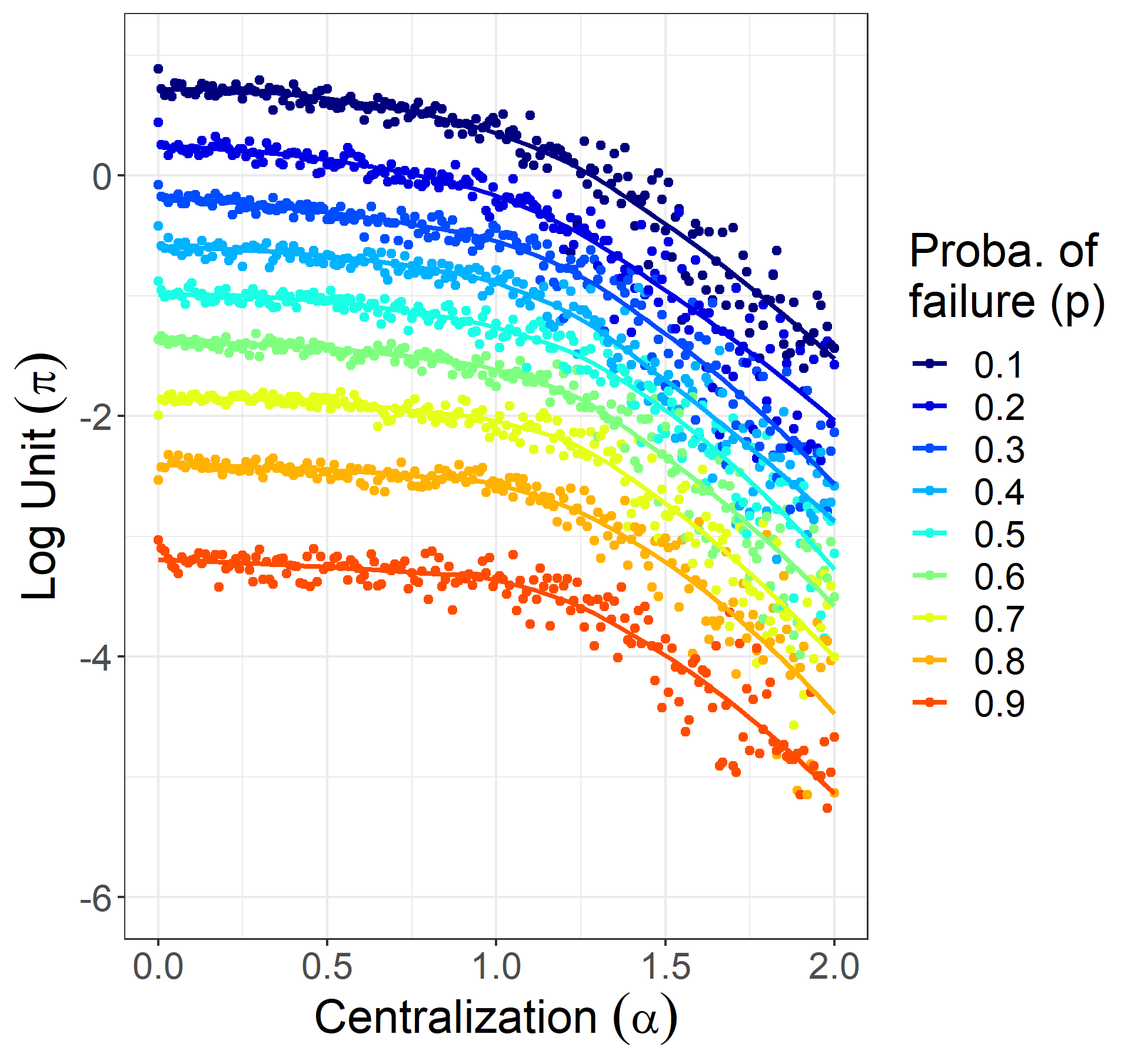}
  \caption{Centralized model}
  \label{productivity_centrality_transitions}
\end{subfigure}
\caption{System productivity and network properties. Productivity is measured in units of the reference level of production in the autarky regime (y-axis). The x-axis shows the network density (left panel) or centralization (right panel). Dots show the results of model simulations. Solid lines show the fitted curve using Polynomial Regression.  Colors indicate distinct values of failure probability. Scale in figure. 
}
\label{productivity_transitions}
\end{figure}

Despite similarities, the results presented in Figures \ref{productivity_sampling} and \ref{collapse_sampling} also show differences between the two network models. These differences are manifested in the way system productivity changes as we vary failure probability. Figure \ref{productivity_transitions} shows system productivity (Log unit) as a function of network density (left) and centralization (right) for various values of individual failure probability (color). The curves exhibit non linear behaviors.


In the density model (left panel in Figure \ref{productivity_transitions}) there seem to be two distinct behaviors depending on the individual failure probability (color). If individual failure probability is below 0.4 (blue), the curves are concave downward and present a maximum value at density values that depend on the failure probability (for example at $\delta=0.02$ for $p=0.2$ or $\delta=0.0175$ for $p=0.2$). The interval in which network density has a positive effect on productivity becomes narrower as individual failure probability increases. Above a failure probability of 0.4, the curves change their behavior and become concave upward (green, yellow and red). In this case, higher density results immediately in a decrease of productivity values regardless of the initial network density. On the other hand, in the network centralization model (right panel in Figure \ref{productivity_transitions}), the curves are always concave downward and monotonically decreasing. In this case, an inflection point that accelerates the decrease of productivity appears when hubs start to gain more importance in the network ($\alpha>1$).



\subsection{Collapse probability and network structure}

\begin{figure}[t!]
\centering
\begin{subfigure}{.5\textwidth}
  \centering
  \includegraphics[width=1.0\linewidth]{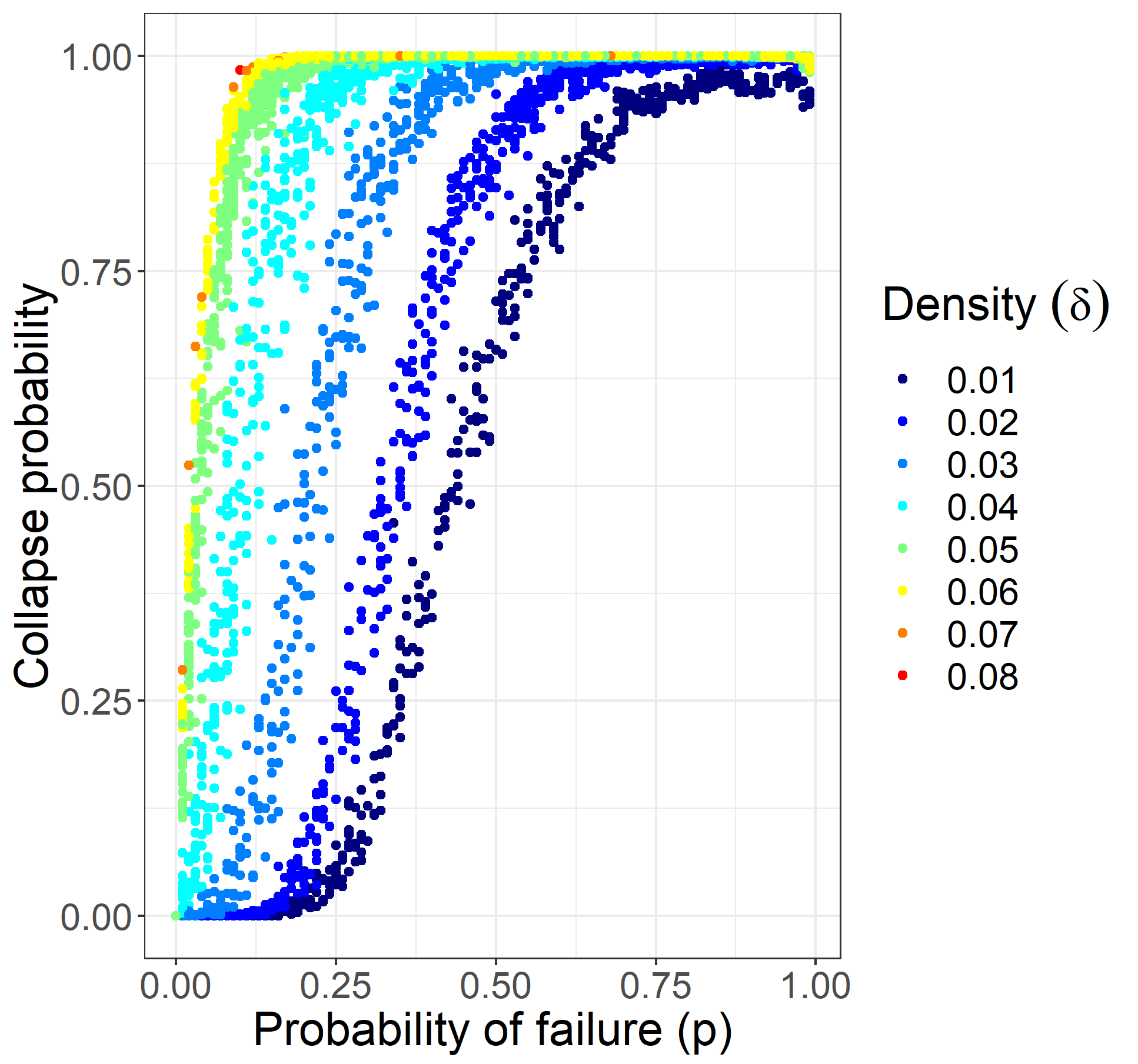}
  \caption{Density model}
  \label{collapse_density_transition_analog}
\end{subfigure}%
\begin{subfigure}{.5\textwidth}
  \centering
  \includegraphics[width=1.0\linewidth]{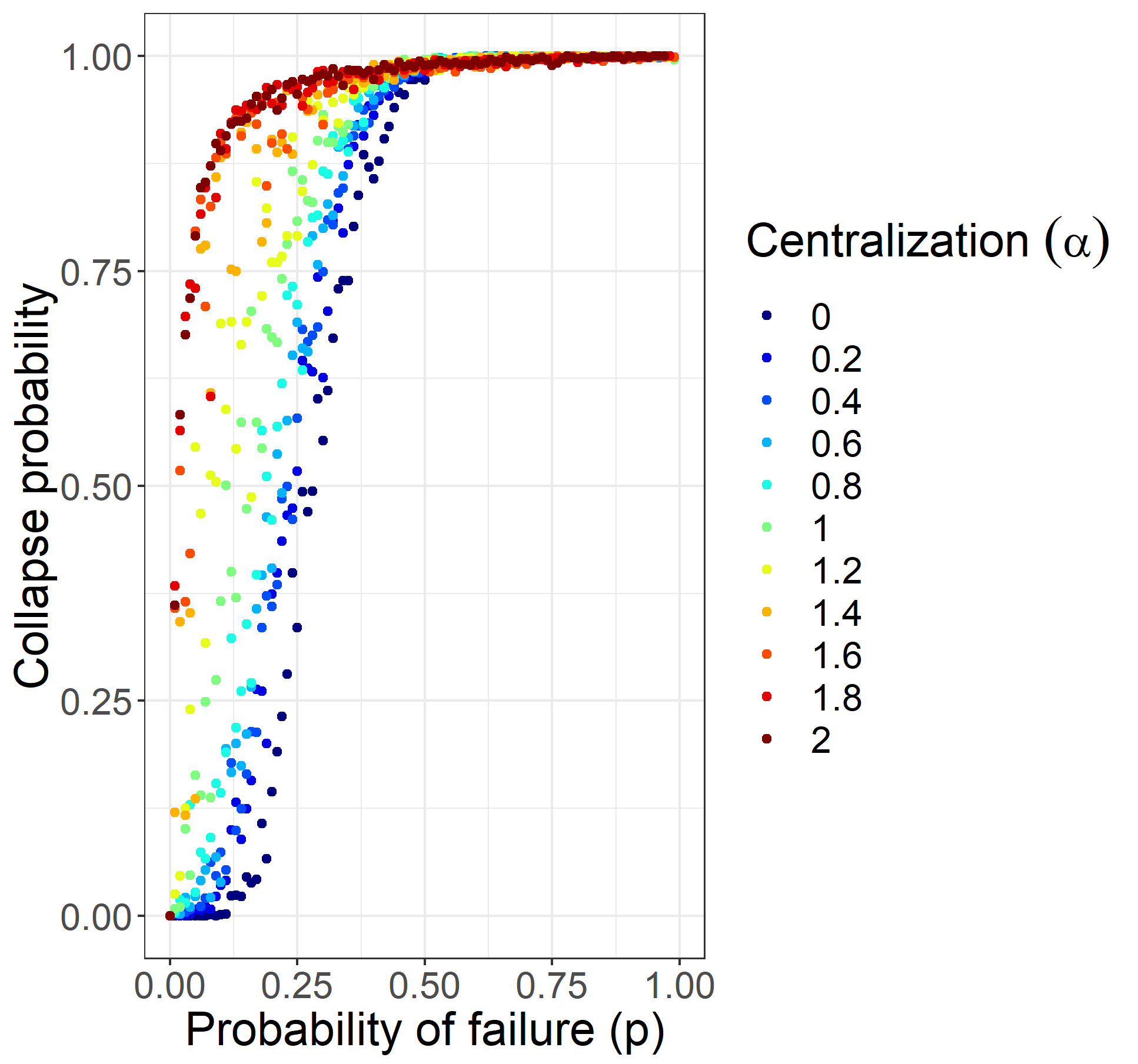}
  \caption{Centralized model}
  \label{collapse_centrality_transition_analog}
\end{subfigure}
\caption{Collapse probability and network properties. The y-axis represents the collapse probability as explained in section \ref{collapse_definition}. The x-axis represents the probability of failure. Dots show the results of model simulations. Colors indicate distinct values of network density (left) and centralization (right). Scale in figure.}
\label{collapse_transition}
\end{figure}

In Figure \ref{collapse_transition}, we analyze the behavior of the collapse probability as a function of individual failure probability by horizontally slicing the surface shown in Figure \ref{collapse_sampling} at different values of density (left panel) or centralization (right panel). 

In both cases, the curves are monotonically increasing, showing that the risk of spreading failure is aligned with both network density and centralization. However, the behavior is non-linear. For small values of failure probability (dark blue curves) there is an interval of network density or centralization in which the system seems robust. In the case of the density model, the robust interval coincides with the location of maximum productivity points shown in Figure \ref{productivity_density_transitions}. As the networks get denser or more centralized (green, yellow and red curves in Figure \ref{collapse_transition}) the extent of the robust interval gets narrower, confirming that the excess of interdependencies increases the fragility of the system. Such decrease occurs more rapidly in the case of centralized networks. \par 

 Using data from the Observatory of Economic Complexity, we apply the model to estimate the vulnerability of international trade networks from 1962 to 2012. Nodes represent countries and edges are present if they have traded on a given year. In the supplement, (Figure \ref{trade} in Section \ref{trade_section}) we show that the  the expansion of interconnections among countries has increased the global sensitivity to failure from 1962 to 2012, with a peak in 2007 just before the last major economic crisis. 
 



\subsection{Universality of collapse}

\begin{figure}[t!]
\centering
\begin{subfigure}{.5\textwidth}
  \centering
  \includegraphics[width=1.0\linewidth]{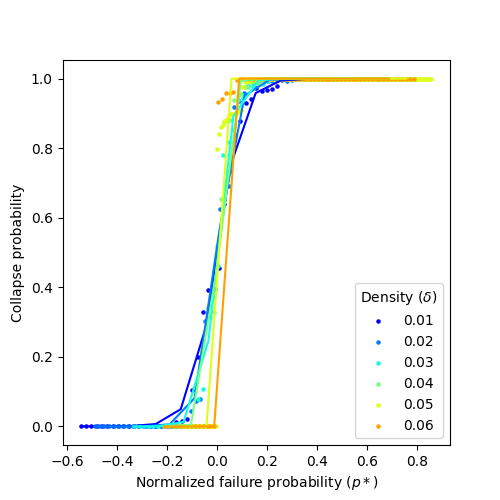}
  \caption{Density model}
  \label{sigmoid_density_reversal}
\end{subfigure}%
\begin{subfigure}{.5\textwidth}
  \centering
  \includegraphics[width=1.0\linewidth]{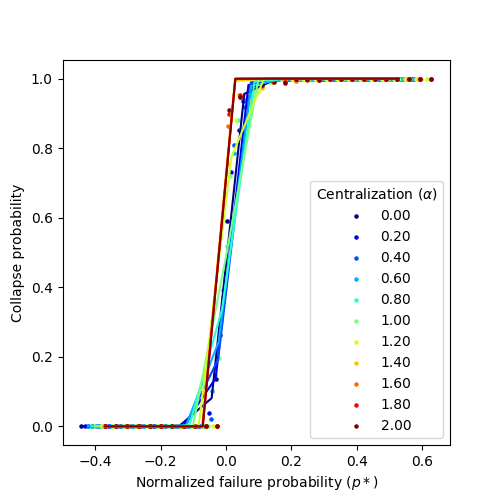}
  \caption{Centralized model}
  \label{sigmoid_centrality_reversal}
\end{subfigure}

\caption{Universal behavior of collapse probability. The left panel shows the results for the density model. The right panel shows the results for the centralized model. Dots represent the resulting collapse probability (y-axis) of model simulations. The solid lines show the fit to the sigmoid function. Colors indicate the respective model parameters (scale inset). The x-axis represents the normalized failure probability ($p*$), after subtracting the location parameter of the sigmoid curve.}
\label{sigmoid}
\end{figure}

The curves in Figure \ref{collapse_transition} suggest that the relationship between collapse and interdependencies (either in the form of density or centralization) can be modeled by a sigmoid function, with the following form: 

\begin{equation}
    f(x) = \frac{1}{1 + e^{-a(x-b)}}
\end{equation}

where $a$ determines the slope of the transition and $b$ the location of the inflection point. In Figure \ref{sigmoid}, we present the results of fitting the sigmoid function to the collapse probability  as a function of failure probability for both network models, at various levels of density (left) and centralization (right) respectively. In order to collapse the curves we normalize the original failure probability ($p$) by subtracting the location parameter of the sigmoid function ($b$), such that $p*=p-b$.


The ubiquity of sigmoidal patterns in the transition to collapse on such different network models and various parameters suggests the existence of a universal behavior. As shown in Figure \ref{collapse_transition}, the sigmoids are present in both types of networks and increasing interdependencies simply moves the inflection point closer to the origin and yields steeper slopes, which indicates higher sensitivity to errors and system fragility. These results indicate that two radically different economic systems, such as centralized and decentralized economies, may fail because of one consistent reason which lies in the dynamics of failure propagation across networks and excess of direct or indirect interdependencies. Like in other complex systems, universalities represent the general structure in which phenomena take place. While individual instances may present different and heterogeneous details, i.e. prices, markets, bureaucracy, etc., there is an underlying structure that is common among them and in which they develop. In order to achieve effective solutions, we must understand and intervene in such structure. Otherwise, there is a risk of spending efforts in designing solutions based on the particularities of each case without considering the relevant variables.

\section{Discussion}

\subsection{Too interconnected to thrive?}

Simulations show that production networks get increasingly sensitive to the propagation of individual failure as their topologies get denser and agents linked to each other. The resulting production may decrease faster than linearly with respect to failure probability. Sparser networks may have a lower productivity but are more resilient to probability of failure. Previous research has investigated the risks of creating agents "too big to fail", or more recently "too central to fail" (Battiston et al. 2012b). In the continuity of this observation, our model emphasizes that without further hypotheses, economic agents may in some situations become too interconnected to thrive. 

This observation from model results, notably drawing from Figure \ref{density_proba} (density model), emphasizes the existence of positive returns to interconnections in system robustness below a given density tipping point. Above such a threshold, returns to interconnections play a negative role. 
Such a mapping of production performance and systemic risk in the sense of global failure appears relevant in tackling efficient asset allocation and minimization of systemic risk as a network optimization problem (Pichler et al., 2018).

These results on returns to interconnections also allow us to replicate in a more general context the findings of Albert, Jeong and Barabási (2000) on error tolerance of scale-free networks. Their results indicate that systems exhibit strong robustness below a level of error of 5\%. This is consistent with the results we obtain in the centralized model ($\alpha=1$) when the probability of individual failure $p=0.05$ (Figure \ref{centrality_proba}). In such networks the collapse probability is low and average production is satisfied. Our analysis thus successfully replicates their findings, while generalizing the study of network robustness to a larger range of organizations and organizing principles. 

\subsection{Risk diversification or containment?}

As noted in the founding work of Schweitzer et al. (2009), traditional economic theory often concludes that dense networks enable risk diversification to counterbalance failure (Allen and Gale, 2000). Risk diversification remains relevant in the context of production, as producers may prefer to protect themselves against the failure of suppliers or trade partners (Bar-Yam, 2010). 
Battiston et al. (2007) identified instead that systemic risk may increase with network coupling strength between nodes in credit chains and production networks during bankruptcy propagation. Contrary to the more general policy implications of Allen and Gale (2000), Battiston et al. (2012c) identified that risk diversification not always reduces systemic risk.

Our model contributes to this debate extending the observations over a larger set of  networks topologies. As illustrated in Figure \ref{density_proba}, our model identifies parameter intervals in which increased density is not detrimental to systemic risk, here denoted as collapse probability of the production system, pointing back to the findings of Allen and Gale (2000). The dark blue region of collapse probability in the density model (Figure \ref{density_proba}), show that the increasing network density through risk diversification, here understood as the creation of additional connections, is not harmful to system robustness if individual probabilities of failure are low. This extent, denoted "density threshold", differs according to network topologies. 

Risk diversification through increased density may be beneficial for the expected productivity in certain intervals of network density and individual failure probabilities (Figure \ref{productivity_density}). However, in environments characterized by higher individual risks, increasing network density may lead to major changes in collapse probability, pushing the system towards unstable situations (red regions in Figure \ref{density_proba}). These results are in line with previous research of Battiston et al. (2012a, 2012c). They show endogenous emergence of systemic risk because of feedback effects resulting from an excess of interdependencies. Our model shows that risk diversification improves global robustness only in an interval of individual risk of failure and outlines that systems may become sensitive too sensitive to individual failure if density and centralization are too high. We show that the transition is not smooth and instead it universally follows a sigmoid behavior. Analogue phase transition process are shown in the model of Lorenz et al. (2009). Our model extends this study with networks displaying centralization (Figure \ref{centrality_proba}). In this case, the collapse probability increases more abruptly and is more sensitive to the risk of individual failure.

\subsection{Policy implications}

Previous literature on systemic risk has raised important suggestions for policy actions. Protection measures have been pointed as necessary through identification of essential nodes (Battiston et al., 2012b). Others opted for a systemic risk tax (Leduc and Thurner, 2017b), in order to make bank networks robust to insolvency cascades, or through an adequate credit default swap market, where CDS assets are taxed according to their contribution to systemic risk (Leduc et al., 2017a). Other scholars have emphasized on the importance of taking networks of interdependencies into account for improving systems' resilience (Buldyrev et al., 2010) in the context of contagion (Marsiglio et al., 2019).

Barriers to contain risk contagion may improve global robustness, whether implemented around a centralized node, or distributed across the decentralized network. Further research on such implementation may shed new light on the impact of safety barriers on different network topologies. However, action in centralized networks cannot be reduced to protection on the central node, and may have less effect than expected, as expressed by Braha and Bar-Yam (2006). 

\section{Conclusion}

In summary, we analyzed the effects of establishing interdependencies among economic agents on arising production complexity and systemic risk. We show that while interdependencies are beneficial for creating more complex products, they also create paths for failure propagation and amplify the fragility of the system--an effect often overlooked in the literature of economic complexity. Our results show that different network topologies, such as dense or centralized networks, show universal patterns of behavior, due to the common dynamics of cascading propagation through direct and indirect connections. Understanding universalities is critical in order to achieve effective solutions beyond the particular characteristics of individual cases. Further research accounting for additional policies such as the enforcement of new types of interdependencies, or application to empirical risk estimation and real economic of financial networks, may contribute to identify opportunities for improving the functioning and complexity of economic systems without compromising their robustness and resilience.

\pagebreak

\section*{Supplementary Material}

\setcounter{equation}{0}
\setcounter{figure}{0}
\setcounter{table}{0}
\setcounter{page}{1}
\setcounter{section}{0}
\renewcommand{\theequation}{S\arabic{equation}}
\renewcommand{\thefigure}{S\arabic{figure}}
\renewcommand{\thesection}{S\arabic{section}}

\section{Model implementation}
\label{implementation}

The model is implemented in the agent-based environment Netlogo (Wilensky, 1999). Agent-based modeling offers a relevant tool to study complex systems, as they emphasize the role of individual interactions, that may be local, heterogeneous and interdependent, and displays flexibility in network generation from the node perspective. As noted Wildemeersch et al. (2016), the behavior of interconnected systems as network systems is too complex to be adequately modeled by traditional tools, suggesting quantitative and simulations methods as possible adequate modelling tools to study network resilience and fragility. Nevertheless, the complexity of the dynamics they describe has created a tough challenge to keep the model tractable and its results understandable. A major criticism addressed to multi-agent models and the use of simulation models, in general, is indeed focused on the lack of tractability of the results and of the model dynamics. Dealing with several parameters, complex interactions and issues of stochasticity and randomness, new tools in simulation exploration and analysis may be required to assess solid understanding of model behavior. 

The OpenMole platform introduced by (Reuillon et al. 2010, 2013) and more specifically its embedded Pattern Space Exploration (PSE) algorithm described in (Ch\'{e}rel et al., 2015) provides a useful tool to improve model understanding by looking at unexpected patterns and exploring the space of outputs generated by the model. This method allows pushing the standards of model validation, commonly done by verifying that the model is able to reproduce the patterns to be explained, to test the validity of the model against unexpected input combinations. The Pattern Space Exploration (PSE) algorithm thus allows to identifying all different output patterns generated by a given range of parameters, which may be useful to understand causality in complex simulation models. It permits to identify regardless of observer's assumption the "areas of interest" of the model, i.e. where variation in output occurs. It finally offers significant benefits in computation efficiency, as areas of little or no variation are not explored by the algorithm, in contrast with a classical sampling task. We thus obtain through the PSE method a more efficient and precise understanding of model areas of interest.

\section{Pattern Space Exploration figures}
\label{pse_figures}

\begin{figure}[h!]
\centering
\begin{subfigure}{.5\textwidth}
  \centering
  \includegraphics[width=1.03\linewidth]{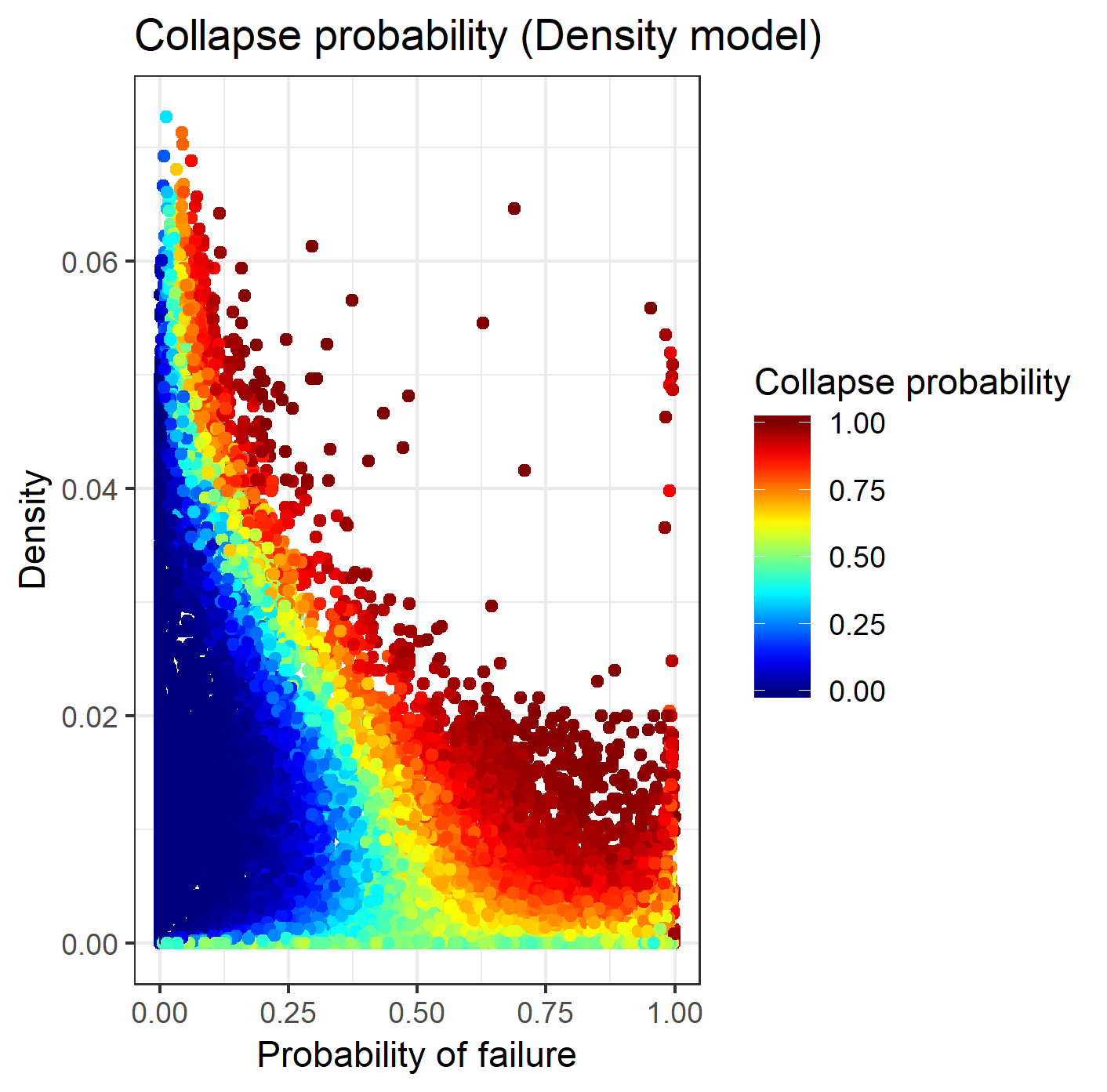}
  \caption{Density model}
\end{subfigure}
\begin{subfigure}{.5\textwidth}
  \centering
  \includegraphics[width=1.03\linewidth]{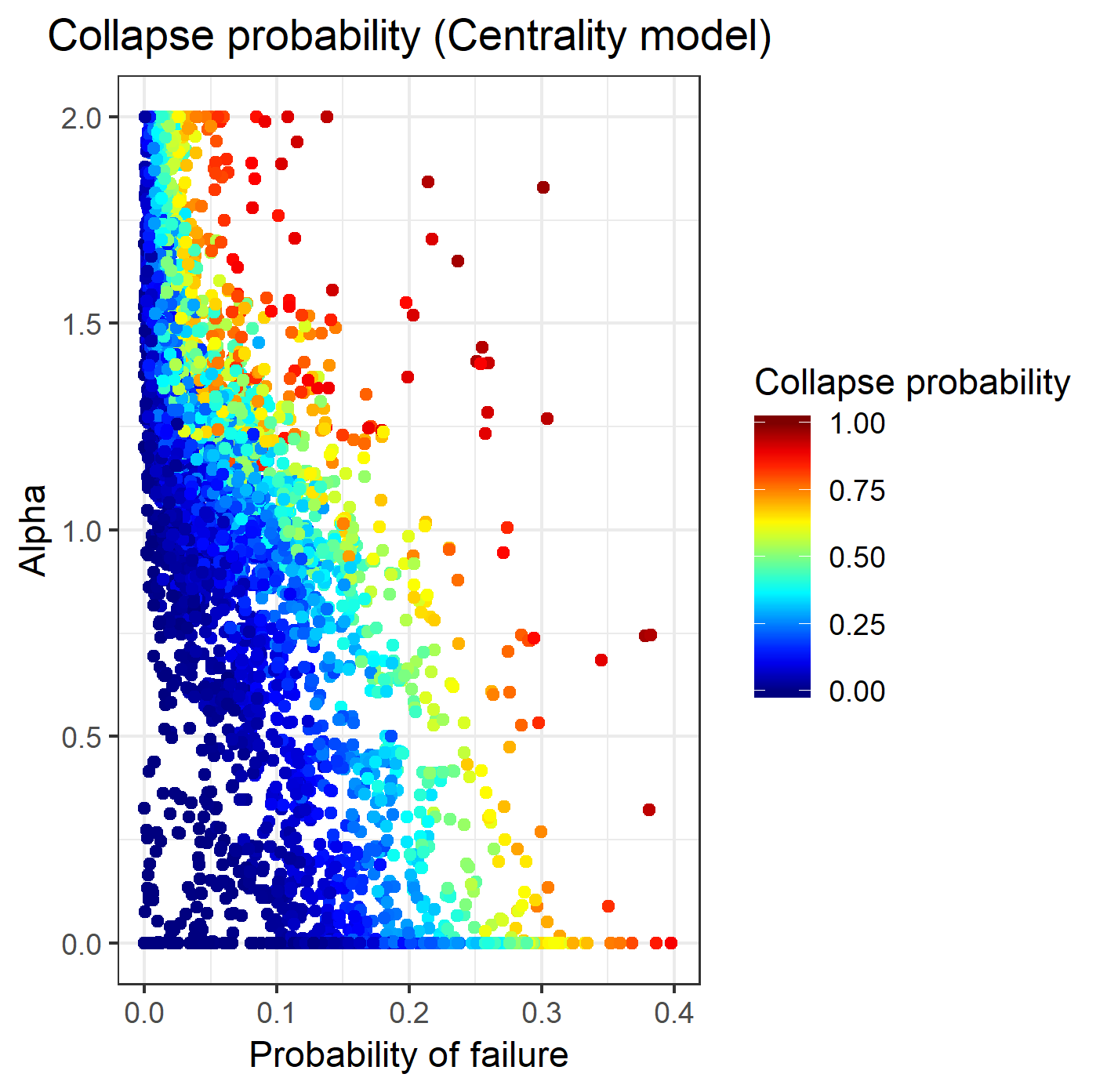}
  \caption{Centralized model}
\end{subfigure}%
\caption{Probability of systemic collapse using Pattern Space Exploration. Color indicates probability of collapse. Scale in figure.}
\label{PSE_collapse}
\end{figure}

\begin{figure}[h!]
\centering
\begin{subfigure}{.5\textwidth}
  \centering
  \includegraphics[width=1.0\linewidth]{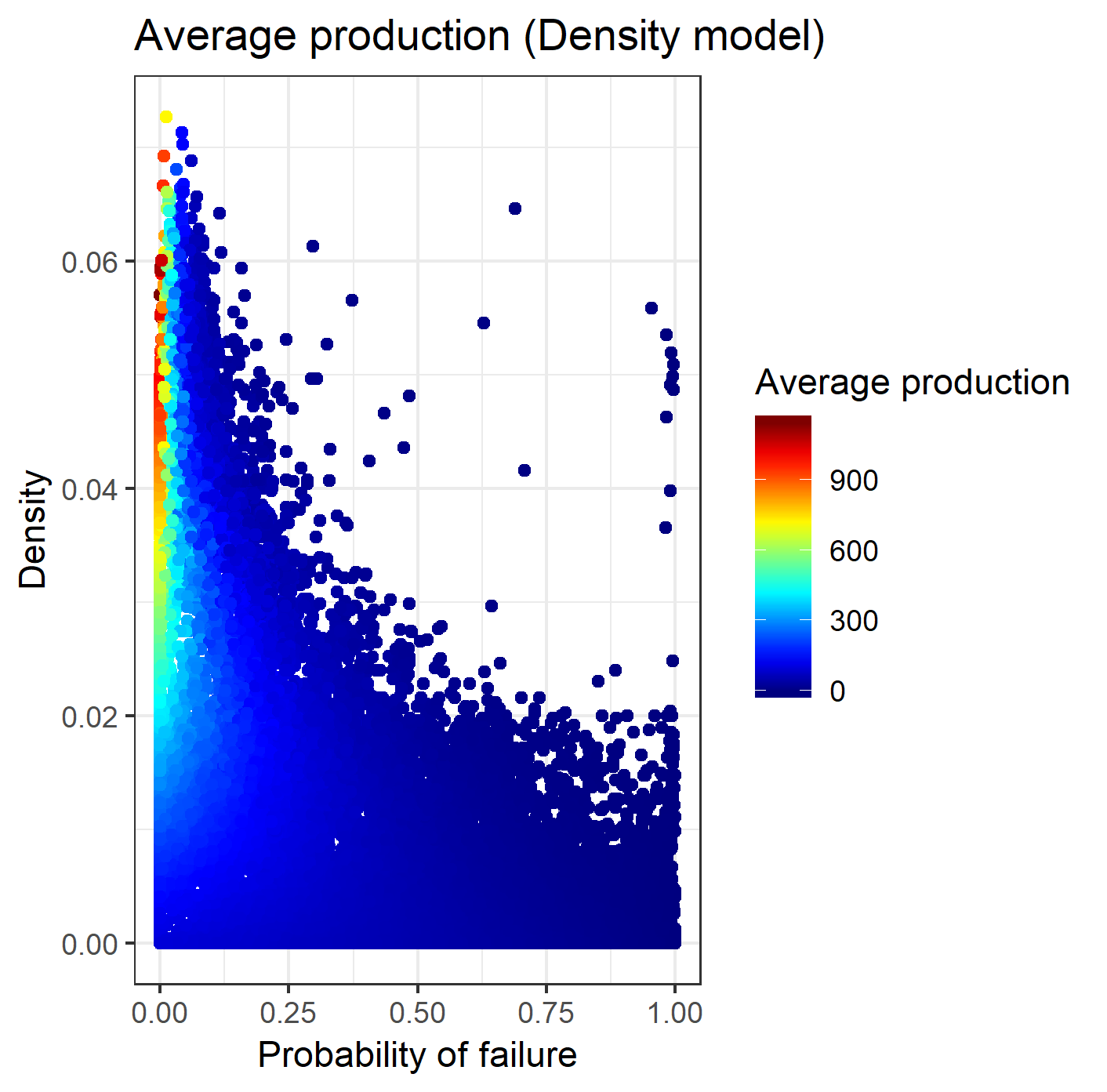}
  \caption{Density model}
\end{subfigure}
\begin{subfigure}{.5\textwidth}
  \centering
  \includegraphics[width=1.0\linewidth]{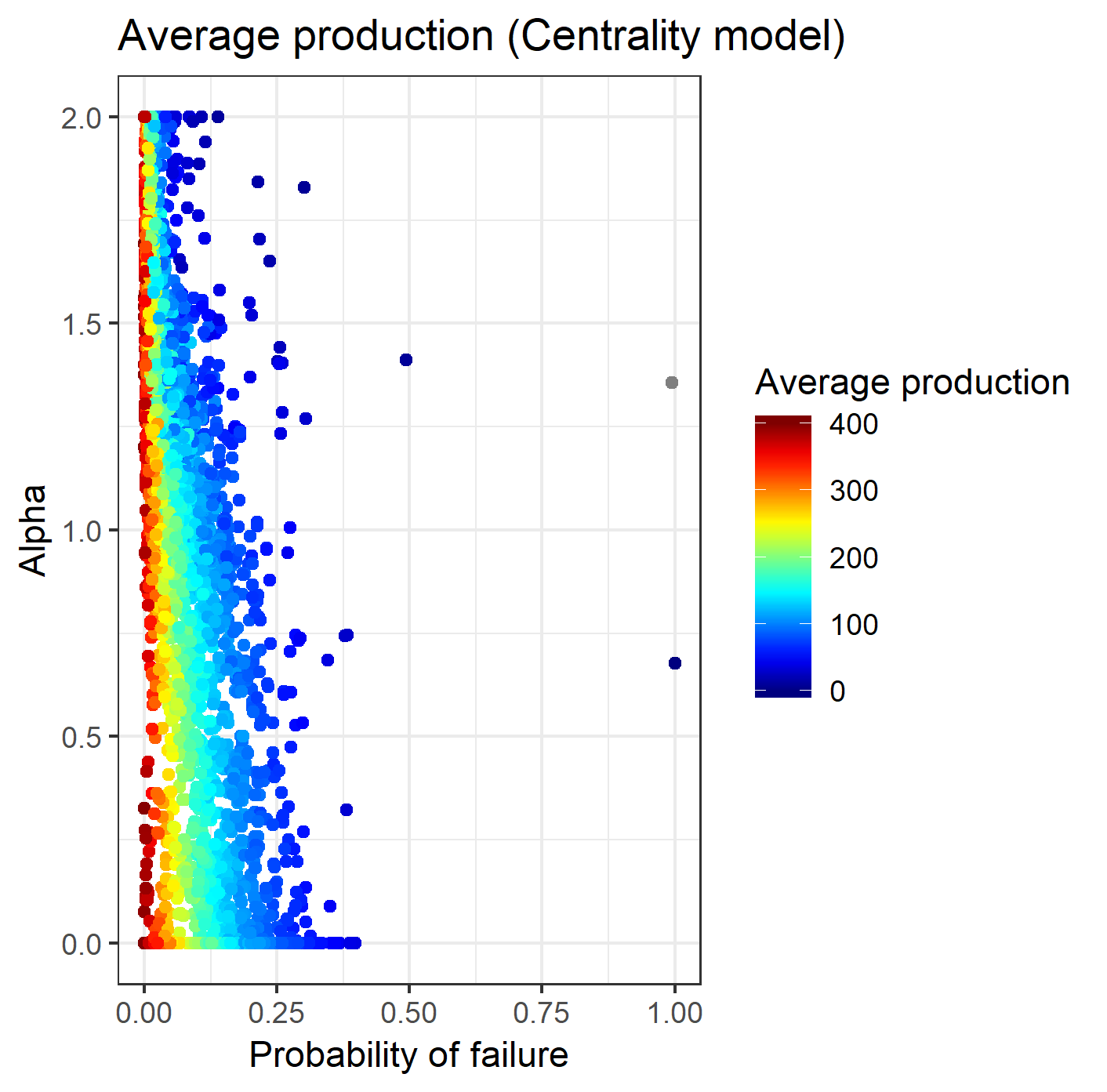}
  \caption{Centralized model}
\end{subfigure}%
\caption{Average production using Pattern Space Exploration. Color indicates average production. Scale in figure.}
\label{PSE_production}
\end{figure}

Figure \ref{PSE_collapse} displays the results of the Pattern Space Exploration (PSE) algorithm implemented in the early steps of the analysis, in order to identify the main dynamics of collapse probability in the model, with respect to individual probability of failure and network structure parameters, i.e. density and centralization. The abrupt impact of higher probability of failures in centralized networks can be observed. The sigmoidal transition in both network structure can also be identified, as well as the low risk (blue) and high risk (red) regions of the parameter space. The white areas in the PSE figures indicate parameter space combinations in which not much variation in output (here for Figure \ref{PSE_collapse} collapse probability) occur. Figure \ref{PSE_production} studies the variations in average aggregate production of the systems in the space of parameters, composed of individual probability of failure and network structure. They outline the relative variations in each topology, and already allow to identify key regions of relative better or worse productivity in the system. The more abrupt transition in the centralization topology towards lower production in the case of excessive interconnections or individual risk is observable.

\section{Trade Networks and Systemic Risk}
\label{trade_section}

\begin{figure}[h!]
    \centering
    \includegraphics[scale = 0.75]{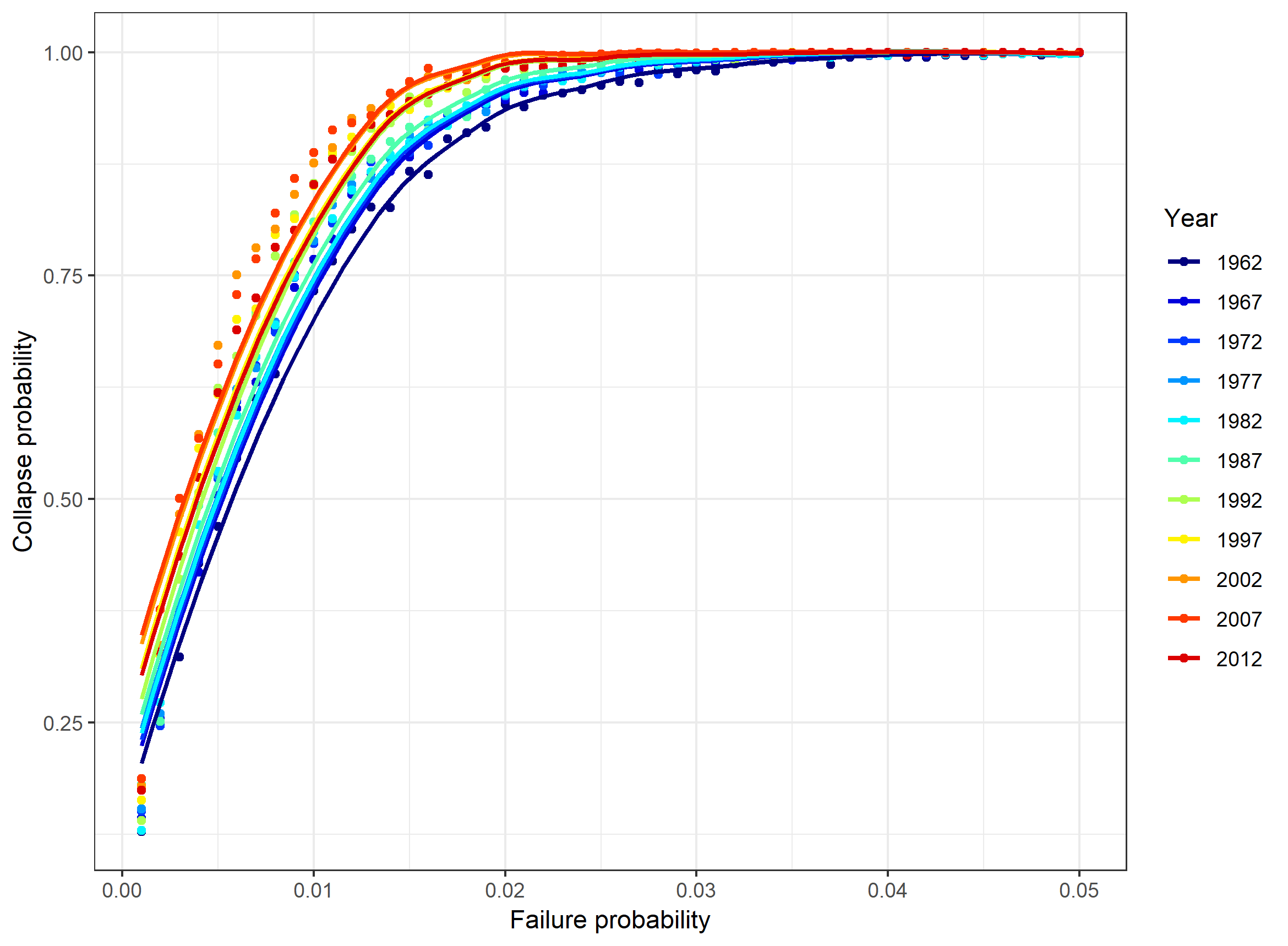}
    \caption{5-year evolution of systemic risk in trade networks computed using our model. Data from the Observatory of Economic Complexity.}
    \label{trade}
\end{figure}

We applied our model to estimate the transition to collapse on global trade networks. The networks have been built with data from the Observatory of Economic Complexity. Nodes represent countries and edges indicate trade at particular years. In Figure \ref{trade} shows that from 1962 to 2012, the systemic risk has increased due to the expansion of interconnections. The figure outlines as well the fragility of trade networks. Certain systemic collapse is attained from a country failure probability of 0.02, i.e. 2\%.

\end{document}